\documentclass[twocolumn,prstab,showpacs,raggedbottom]{revtex4}

\usepackage{graphicx}
\usepackage{amsmath}
\usepackage{subfigure}

\begin{document}

	\title{Constraints on photon pulse duration from longitudinal electron beam diagnostics at a soft X-ray free-electron laser}
	
	\author{C. Behrens$^\mathrm{1}$, N. Gerasimova$^\mathrm{1}$, Ch.~Gerth$^\mathrm{1}$, B. Schmidt$^\mathrm{1}$, E.A.~Schneidmiller$^\mathrm{1}$, S. Serkez$^\mathrm{1,2}$, S. Wesch$^\mathrm{1}$, and M.V. Yurkov$^\mathrm{1}$}

	\affiliation{$^\mathrm{1}$ Deutsches Elektronen-Synchrotron DESY, Notkestr.\,85, 22607 Hamburg, Germany\\
	$^\mathrm{2}$ Ivan Franko National University of Lviv, 8 Kyryla i Mefodiya Street, 79005 Lviv, Ukraine}

	\date{\today}				
	
	\begin{abstract}
	The successful operation of X-ray free-electron lasers (FELs), like the Linac Coherent Light Source or the Free-Electron Laser in Hamburg (FLASH), makes unprecedented research on matter at atomic length and ultrafast time scales possible. However, in order to take advantage of these unique light sources and to meet the strict requirements of many experiments in photon science, FEL photon pulse durations need to be known and tunable. This can be achieved by controlling the FEL driving electron beams, and high-resolution longitudinal electron beam diagnostics can be utilized to provide constraints on the expected FEL photon pulse durations. In this paper, we present comparative measurements of soft X-ray pulse durations and electron bunch lengths at FLASH. The soft X-ray pulse durations were measured by FEL radiation pulse energy statistics and compared to electron bunch lengths determined by frequency-domain spectroscopy of coherent transition radiation in the terahertz range and time-domain longitudinal phase space measurements. The experimental results, theoretical considerations, and simulations show that high-resolution longitudinal electron beam diagnostics provide reasonable constraints on the expected FEL photon pulse durations. In addition, we demonstrated the generation of soft X-ray pulses with durations below 50\,fs (FWHM) after the implementation of the new uniform electron bunch compression scheme used at FLASH.
	\end{abstract}


	\maketitle

	\section{Introduction}\label{sec:Introduction}
	Since the first operation of a laser in 1960~\cite{laser}, a tremendous number of new experimental techniques became possible with continuously changing and growing requirements on the laser systems, e.g., higher spectral brightness, shorter wavelengths, or photon pulse durations with simultaneous tunability. High-gain free-electron lasers (FELs) meet many of theses requirements, and the successful operation of X-ray FELs like the Free-Electron Laser in Hamburg (FLASH)~\cite{FLASHAnature}, the Linac Coherent Light Source (LCLS)~\cite{LCLSnature2}, or the SPring-8 Angstrom Compact Free Electron Laser~\cite{SACLAnature} make unprecedented research on matter at atomic length scales possible~\cite{LCLSnature1}. The demonstration of FEL photon pulse durations in the femtosecond range (e.g., Refs.~\cite{FLASH1,SD}) has further extended the capabilities to research of dynamical processes at ultrafast time scales (see, e.g., Ref.~\cite{Dynamics}) with ongoing demands on the generation and control of ultrashort photon pulses. 

	In recent years, several methods to control FEL photon pulse durations have been proposed by manipulating and controlling the FEL driving electron bunches. The low-charge operation at LCLS~\cite{Ding} demonstrated electron bunch lengths below 10\,fs~\cite{Aline2}, and the same strategy of low-charge operation is planned for the European XFEL and FLASH~\cite{Igor}. Other methods, with additional prospects of generating photon pulses in the attosecond range, make use of electron bunch manipulation with conventional quantum lasers (e.g., Refs.~\cite{SSY1,zho,chirp2,Xiang,SSY2}) or by selectively spoiling the transverse emittance of the electron beam~\cite{Spoiler}. However, reliable operation of theses methods requires capabilities to diagnose the FEL photon pulse shapes and durations with high accuracy, which is a tremendous challenge and an active field of research.
	
	 First single-shot measurements of FEL photon pulse durations with femtosecond accuracy have recently been demonstrated by terahertz-field driven streaking experiments in the time-domain~\cite{THZ_streak2,THZ_streak1}, and the statistical and spectral properties of FEL radiation emitted in the exponential gain regime allow to measure the mean photon pulse duration (e.g., Refs.~\cite{stat-oc,book,SSY3,lcls-fel2010}). Other proposed methods make use of the FEL induced slice energy loss in order to measure ``replicas'' of the FEL photon pulses (see Refs.~\cite{xtcav,oa}). As is presented in this paper, high-resolution longitudinal electron beam diagnostics in standard configurations, i.e., as commonly used at present X-ray FELs, can provide reasonable and complementary constraints on the expected FEL photon pulse durations.

	In this paper, we compare measurements of electron bunch lengths with corresponding soft X-ray pulse durations at FLASH, and discuss their relationship by theoretical considerations and simulations. The experimental setup and applied methods are described in Sec.~\ref{sec:Setup}, and in Sec.~\ref{sec:Results} we discuss the experimental results for FEL operation with different bunch charges, which corresponds to different electron bunch lengths and soft X-ray pulse durations. The final results are compared in Sec.~\ref{sec:Comparisons}, and the summary and conclusions are presented in Sec.~\ref{sec:Summary}.
	
	\section{Experimental setup and methods}\label{sec:Setup}
	\begin{figure*}[htb]
		\centering
		\includegraphics[width=1\linewidth]{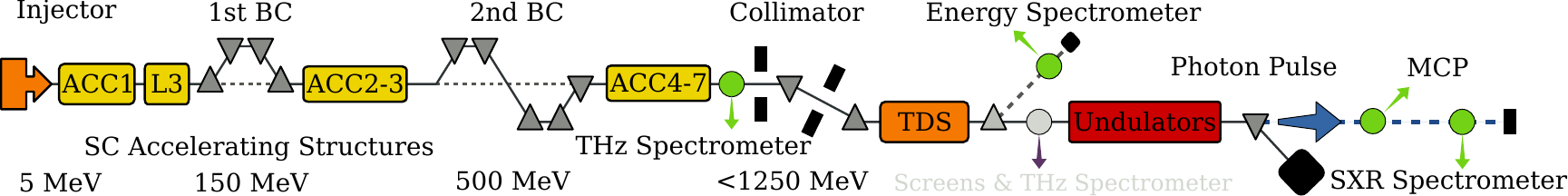} 
		\caption{Layout of the Free-Electron Laser in Hamburg (FLASH) with its superconducting (SC) accelerating structures (ACC), the two magnetic bunch compressor (BC) chicanes, and the new third-harmonic rf linearizer system (L3). The positions of the diagnostics used for photon pulse duration and electron bunch length measurements are indicated by green dots and arrows. }
		\label{fig:FLASH_1}
	\end{figure*}
	The measurements presented in this paper were carried out at FLASH, which is a self-amplified spontaneous emission (SASE) FEL~\cite{kon} for extreme-ultraviolet and soft X-ray radiation, driven by a superconducting radio-frequency (rf) linear accelerator~\cite{FLASHAnature}. The schematic layout of FLASH is depicted in Fig.~\ref{fig:FLASH_1}, showing the injector, which is based on a laser-driven normal conducting rf gun, the superconducting accelerating structures, two magnetic bunch compressor chicanes, and the undulator magnet system. The positions of the diagnostics used for FEL photon pulse duration and electron bunch length measurements are indicated by green dots and arrows. 

	In 2010, FLASH was upgraded~\cite{flash} to a maximum beam energy of 1.25\,GeV, allowing for the generation of soft X-rays below 4.4\,nm (water window) with its fundamental. One of the major upgrades was the installation of the third-harmonic rf system, which is dedicated for the linearization of the longitudinal phase space upstream of the first bunch compressor~\cite{lin1,lin2}. Before this upgrade, FLASH was operated with non-uniformly compressed electron bunches with a short leading spike~\cite{ls,Roehrs}. The corresponding FEL photon pulses had short durations but low pulse energies due to the fact that only a small fraction of the bunch (the short leading spike) contributed to lasing~\cite{FLASHAnature, FLASH1}. In Sec.~\ref{sec:Results}, it is shown that the third-harmonic rf linearizer system permits more uniform bunch compression with higher FEL photon pulse energies and simultaneous tunability of the pulse durations.
		
	In the following, we describe the experimental setup and applied methods to measure the FEL photon pulse durations and electron bunch lengths. The photon pulse durations were estimated by the statistics of FEL radiation pulse energies measured with a micro-channel plate based detector. The electron bunch lengths were measured in the time-domain by using a transverse deflecting rf structure in combination with a magnetic energy spectrometer and in the frequency-domain by spectroscopy of coherent transition radiation in the terahertz range.

		\subsection{Time-domain longitudinal phase space diagnostics for electron beams}\label{subsec:LOLA}
		The time-domain electron bunch length measurements were done by using a transverse deflecting rf structure (TDS)~\cite{LOLA} in combination with a magnetic energy spectrometer. Transverse deflecting rf structures are widely used for electron bunch length and longitudinal profile measurements at present FELs, and provide high-resolution single-shot diagnostics (see, e.g., Ref.~\cite{Kick}). In combination with standard techniques for transverse emittance measurements, the operation of TDSs makes time-resolved emittance, i.e., slice emittance, measurements possible. The complementary use of an energy spectrometer allows direct longitudinal phase space and slice energy spread measurements (e.g., Refs.~\cite{Roehrs,Filippetto,LOLA1,LOLA2}). Recently, a TDS in combination with a magnetic energy spectrometer downstream of FEL undulators have been proposed to measure the FEL-induced slice energy spread for temporal X-ray pulse characterization~\cite{xtcav}.

		At FLASH, a LOLA-type~\cite{LOLA} TDS was successfully operated upstream of the energy collimator~\cite{Roehrs} before it was moved close to the undulators during the FLASH upgrade in 2010~\cite{flash}. The TDS has been integrated in a dedicated setup for measurements of the longitudinal phase space~\cite{LOLA1,LOLA2}. As depicted in Fig.~\ref{fig:FLASH_1}, the TDS can either be operated in combination with the dispersive energy spectrometer or by using off-axis screens in the non-dispersive main beamline during FEL operation. In both cases, the screen stations are equipped with different imaging screens and a camera system with motorized optics. A fast kicker magnet (not shown in Fig.~\ref{fig:FLASH_1}) can operate the off-axis screens for TDS measurements and frequency-domain electron bunch length diagnostics by terahertz spectroscopy (see Sec.~\ref{subsec:kk}) in parallel. Technical details and performance measurements on the new measurement setup can be found in Refs.~\cite{LOLA1,LOLA2}, and technical information about the TDS at FLASH and detailed descriptions of time-domain electron bunch diagnostics using a TDS can be found in Refs.~\cite{Roehrs,IES2}. Here, we describe only the basic principles of longitudinal phase space diagnostics required throughout this paper.

		The vertical betatron motion of an electron passing a TDS around rf phase zero-crossing is given by~\cite{Roehrs,IES2}
		\begin{equation}
			y_\pm(s) =  y_0(s) + C_y(s,s_0) c^{-1}z  \pm S_y(s,s_0)c^{-1}z 
		\label{eq:motion}
		\end{equation} with the linear correlation $C_y(s,s_0)$ and shear function
		\begin{equation}
			S_y(s,s_0) = R_{34}K_y= \sqrt{\beta_y(s)\beta_y(s_0)}\mathrm{sin}(\Delta\phi_y) \frac{e \omega V_y}{E}\,,
		\label{eq:beta}
		\end{equation} where $R_{34}=\sqrt{\beta_y(s)\beta_y(s_0)}\mathrm{sin}(\Delta\phi_y)$ is the angular-to-spatial element of the vertical beam transfer matrix from the TDS at $s_0$ to any position $s$, $\beta_y$ is the vertical beta function, $\Delta\phi_y$ is the vertical phase advance between $s_0$ and $s$, and $y_0$ describes the intrinsic offset. The expression $K_y = e \omega V_y/E$ is the vertical kick strength with the peak deflection voltage $V_y$ in the TDS, $c$ is the speed of light in vacuum, $e$ is the elementary charge, $E$ is the electron energy, $z$ is the longitudinal position of the electron relative to the zero-crossing rf phase, and $\omega/(2\pi)$ is the operating rf frequency. The longitudinal-to-vertical correlation $C_y$ in Eq.~(\ref{eq:motion}) is independent of the TDS operation and can exist intrinsically due to time-dependent kicks generated from collective effects such as coherent synchrotron radiation or wakefields. This linear correlation may lead to systematical errors in electron bunch length measurements, which, however, can be removed by performing measurements at two zero-crossing TDS rf phases shifted by 180\,deg, i.e., with $\pm S_y$ (cf. Eq.~(\ref{eq:motion})).

		The expression in Eq.~(\ref{eq:motion}) shows a linear mapping from the longitudinal to the vertical coordinate and allows longitudinal electron bunch profile measurements by means of transverse beam diagnostics using imaging screens or wire-scanners. The vertical shear function $S_y$ determines the slope of this mapping and can be calibrated by measuring the vertical centroid offset of the bunch as a function of the TDS rf phase. The electron bunch current is given by the calibrated and normalized longitudinal bunch profile multiplied by the measured electron bunch charge. In the following, the longitudinal coordinate $z$ is expressed by a time coordinate via $t=-z/c$ and the bunch length by the corresponding bunch duration. The latter is expressed by the root mean square (r.m.s.) value $\Sigma_{t,\mathrm{e}} = |C_y\pm S_y|^{-1} (\sigma_{y\pm}^2 - \sigma_{y0}^2)^{1/2}$, where $\sigma_{y\pm}$ is the measured vertical r.m.s.~beam size during TDS operation corresponding to $\pm S_y$, respectively. The r.m.s.~bunch duration $\Sigma_{t,\mathrm{e}}$ describes a quadratic equation for $\sigma_{y\pm}^2$, and by performing measurements of the vertical r.m.s.~beam size at $\pm S_y$ and $S_y=0$ (TDS switched off), the longitudinal-to-vertical correlation $C_y$ and the intrinsic vertical r.m.s.~beam size $\sigma_{y0}$ can be determined. These parameters limit the r.m.s.~time resolution to $\mathcal{R}_{t,\mathrm{e}}=\sigma_{y0}/|C_y\pm S_y|$ (cf. Refs.~\cite{Roehrs,IES2}).

		By combining the operation of a TDS with an energy spectrometer and using imaging screens to get two-dimensional transverse beam profiles, longitudinal phase space (energy versus time) measurements with single-shot capability can be achieved. The simplest magnetic energy spectrometer consists of a dispersive beamline downstream of a dipole magnet. The corresponding horizontal betatron motion, which is perpendicular to the shearing plane of the TDS, is given by (e.g., Refs.~\cite{Roehrs,IES2})
		\begin{equation}
			x(s) =  x_0(s) + D_x(s,s_0)\delta
		\label{eq:disp}
		\end{equation} with the horizontal dispersion $D_x$ and the relative momentum deviation $\delta = \Delta p/p$. The magnetic energy spectrometer used at FLASH has a nominal horizontal dispersion of about 750\,mm at the position of the screen station~\cite{LOLA2}. For relativistic electron beam energies with a Lorentz factor of $\gamma \gg 1$, which is the case throughout this paper, the electron beam energy is given by $E\approx pc$, and $\delta$ can be described as the relative energy deviation. Then the expression in Eq.~(\ref{eq:disp}) represents a linear mapping between the relative energy deviation and the horizontal coordinate, where $D_x$ determines the slope of the mapping. The dispersion $D_x$ can be calibrated by measuring the horizontal centroid offset of the bunch as a function of the energy deviation. The r.m.s.~resolution of the relative energy deviation is defined as $\mathcal{R}_{\delta,\mathrm{e}}=\sigma_{x0}/D_x$ (cf. time resolution $\mathcal{R}_{t,\mathrm{e}}$), but here, the intrinsic beam size, and correspondingly the energy resolution cannot simply be measured by switching the dipole magnet off. A possibility to estimate the relative energy resolution is given by the deviations in the measurement of a well-known energy spread (see Refs.~\cite{Roehrs,LOLA2} for more details).

		\subsection{Frequency-domain terahertz spectroscopy of coherent transition radiation}\label{subsec:kk}
\begin{figure}[htb]
		\centering
		\subfigure[~Spectrometer layout with multiple stages (top view).]{\includegraphics[width=0.9\linewidth]{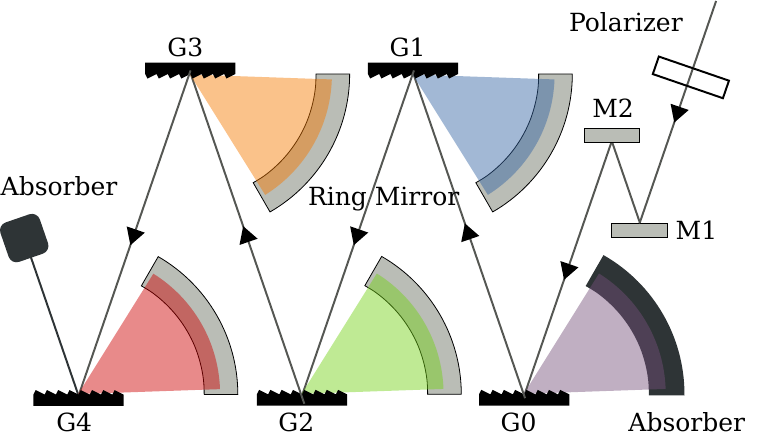} \label{fig:THz_2_a}}
		\subfigure[~One stage with grating, ring mirror, and detector.]{\includegraphics[width=0.8\linewidth]{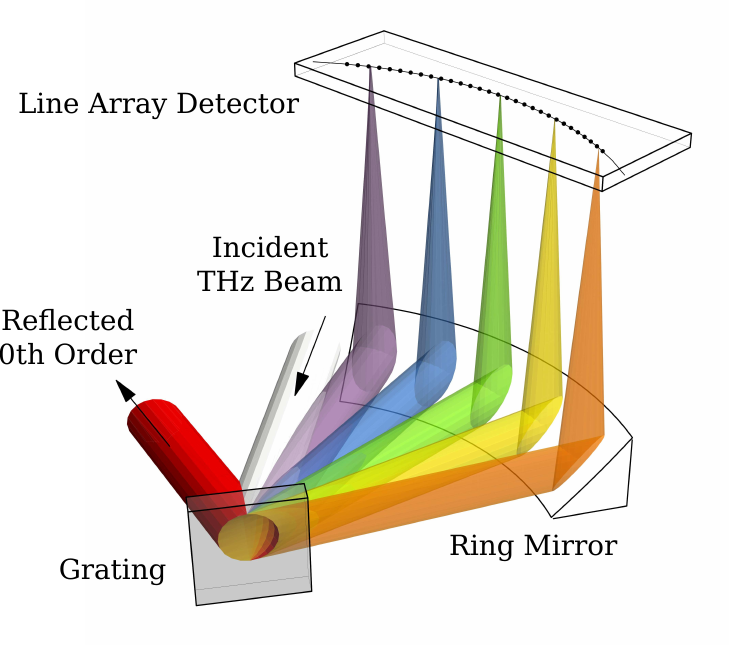} \label{fig:THz_2_b}}
		\caption{Principle of the multi-channel THz and infrared spectrometer based on five stages of blazed reflection gratings in combination with focusing ring mirrors. The schematic layout is depicted in (a), and (b) shows one stage exemplarily (see Refs.~\cite{THz,Wesch1} for technical details). The fast readout of the pyroelectric detector arrays allows single-shot spectroscopy. }					
		\label{fig:THz_2}
		\end{figure}

		The frequency-domain longitudinal electron bunch profile measurements were carried out by spectroscopy of coherent transition radiation using a multi-channel terahertz (THz) spectrometer. The schematic and basic layout of the THz spectrometer is shown in Fig.~\ref{fig:THz_2}. The design is based on five stages of blazed reflection gratings in combination with focusing ring mirrors. Fast readout of the pyroelectric detector arrays allows single-shot diagnostics. The technical details about the focusing ring mirror design and the spectrometer principle can be found in Ref.~\cite{THz}, and technical details and performance measurements on the particular THz spectrometer used throughout the measurements presented in this paper can be found in Ref.~\cite{Wesch1}. At FLASH, two identical THz spectrometers exist, one upstream of the energy collimator (see Fig.~\ref{fig:FLASH_1}), which was used for the spectroscopic measurements presented in this paper, and one close to the undulators, which is currently being commissioned.

		The basic principle of spectroscopy for longitudinal electron bunch profile diagnostics relies on the spectral intensity of transition radiation (diffraction or synchrotron radiation are also suitable) emitted by an electron bunch with $N$ electrons which is given by~\cite{Wesch1,Lai1,Piot,Grimm}
		\begin{equation}
			\frac{d\mathcal{I}}{d\lambda} =\frac{d\mathcal{I}_1}{d\lambda} \left(N+(N-1)N |F(\lambda)|^2\right)\,,
		\label{eq:spec}
		\end{equation}
		where $d\mathcal{I}_1/d\lambda$ is the spectral intensity of a single electron, and $F(\lambda)$ is the longitudinal form factor of the bunch as function of the wavelength $\lambda$, which can be expressed by the Fourier transform of the normalized longitudinal charge density $\rho_N(z)$, i.e., the electron bunch current, as
		\begin{equation}
			F(\lambda)= \int_{-\infty}^{\infty}{\rho_N(z)\mathrm{e}^{-i 2\pi  z/\lambda } dz}\,.
		\label{eq:form}
		\end{equation}
		In Eq.~(\ref{eq:spec}) we have made use of the fact that radiation from relativistic electrons is confined to small angles, hence we have neglected transverse beam size effects~\cite{Wesch1}. Inverse Fourier transform permits the reconstruction of the longitudinal electron bunch profile, i.e. the bunch current $I(z)=Q \rho_N(z)$ with the bunch charge $Q$, when the complex form factor $F(\lambda)=|F(\lambda)|\mathrm{e}^{i\Phi(\lambda)}$ is known. 
		 
		Only the modulus of the longitudinal form factor $|F(\lambda)|$ can be determined experimentally (see Eq.~(\ref{eq:spec})), and the phase information $\Phi(\lambda)$ remains unknown. However, a phase retrieval can be achieved by applying the Kramers-Kronig relations (e.g., Refs.~\cite{Lai1,Piot,Grimm}), which connect the real and imaginary parts of a complex function. Taking the logarithm of the complex form factor $\ln F= \ln|F| +i \Phi$ and applying a Hilbert transform yield
		\begin{align}
			\Phi(\omega)= -& \frac{1}{\pi}\mathcal{P}\int_{-\infty}^{\infty}{\frac{\ln|F(\omega')|}{\omega'-\omega}d\omega'} + \Phi_{\mathrm{B}}(\omega)\nonumber \\
			= - & \frac{2 \omega}{\pi }\mathcal{P}\int_{0}^{\infty}{\frac{\ln|F(\omega')|}{\omega'^2-\omega^2}d\omega'} + \Phi_{\mathrm{B}}(\omega)\,,
		\label{eq:kk1}
		\end{align}
		where $\Phi_{\mathrm{B}}$ is the Blaschke phase, $\mathcal{P}$ denotes the Cauchy principal value, and $\omega=2\pi c/\lambda$.	The Blaschke phase cannot be determined from the modulus of the form factor and is omitted ($\Phi_{\mathrm{B}}(\omega)\equiv0$) in the following. A profile reconstruction with the remaining minimal phase gives the most compact profile compatible with the measured form factor $|F|$. In general, this is a good approximation, as demonstrated in Sec.~\ref{subsec:ctr}, and uncertainties due to measurement errors may have larger impact (see, e.g., Refs.~\cite{Lai1,Grimm}). Changing to wavelengths and removing the singularity at $\omega=\omega'$ in Eq.~(\ref{eq:kk1}) result in~\cite{Lai1,Grimm}
		\begin{equation}
			\Phi(\lambda)= \frac{2}{\pi\lambda}\int_{0}^{\infty}{\frac{\ln(|F(\lambda')|/|F(\lambda)|)}{1-(\lambda'/\lambda)^2}d\lambda'}\,.
		\label{eq:kk2}
		\end{equation}
		The inverse Fourier transform of Eq.~(\ref{eq:form}) can be given by
		\begin{equation}
			\rho_N(z)=-2\int_{0}^{\infty}{|F(\lambda')|\mathrm{cos}\bigl(2\pi  z/\lambda' - \Phi(\lambda')\bigl) \frac{d\lambda'}{\lambda'^2}}
		\label{eq:kk3}
		\end{equation}
		when considering only positive wavelengths ($\lambda>0$). The Eqs.~(\ref{eq:kk2}) and~(\ref{eq:kk3}) are the fundamental expressions for longitudinal electron bunch profile reconstruction from spectroscopic measurements and are used later in Sec.~\ref{subsec:ctr}.

	\subsection{Statistical fluctuations of the radiation pulse energy in SASE FELs}\label{subsec:FEL}
	
\begin{figure}[t]
			\centering
			\subfigure[~Absolute values.]{\includegraphics[width=0.9\linewidth]{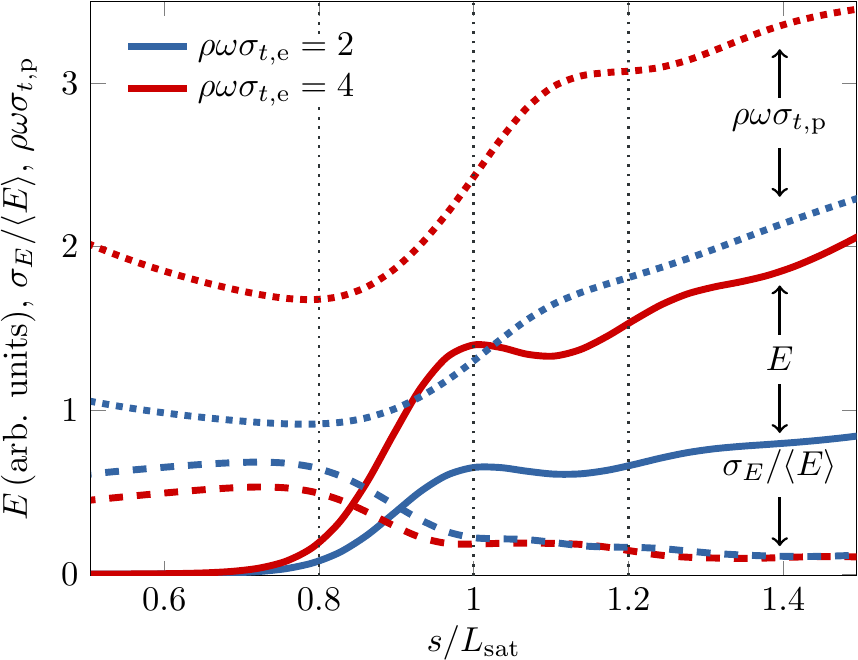} \label{fig:FEL_3_a}}
			\subfigure[~Normalized values.]{\includegraphics[width=0.9\linewidth]{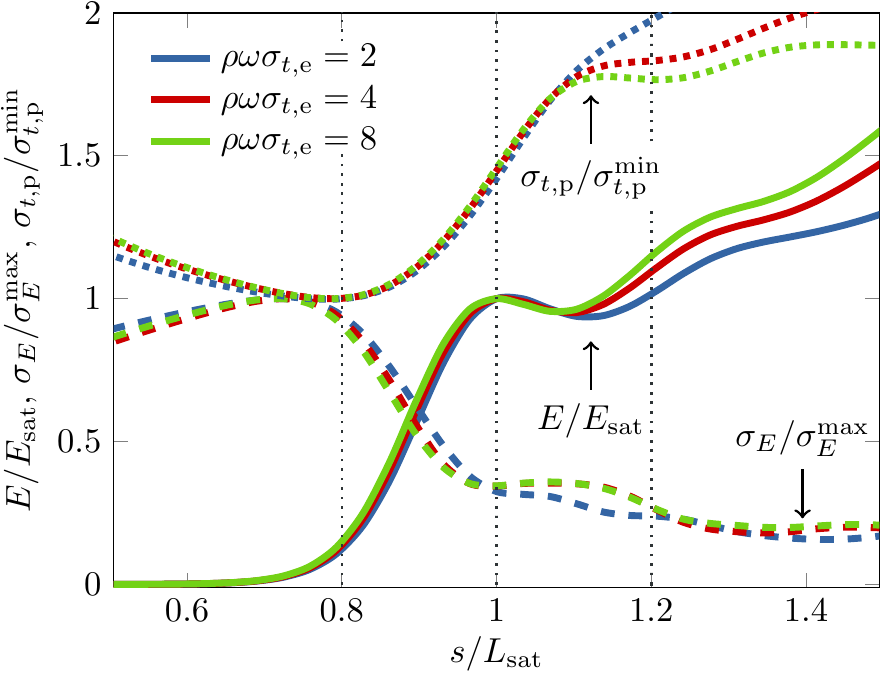} \label{fig:FEL_3_b}}
			\caption{Simulations on the evolution of FEL related parameters versus normalized undulator length $s/L_{\mathrm{sat}}$: (a) Evolution of the energy $E$ in the radiation pulse (solid lines), the fluctuations of the radiation pulse energy $\sigma_{E}/\langle E \rangle$ (dashed lines), and the normalized r.m.s.~FEL radiation pulse duration $\rho\omega\sigma_{t,\mathrm{p}}$ (dotted lines) for normalized r.m.s.~electron bunch lengths of $\rho\omega\sigma_{t,\mathrm{e}} = 2$ and $4$ (blue and red), and (b) the normalized values of $E/E_{\mathrm{sat}}$, $\sigma_{\mathrm{E}}/\sigma_{\mathrm{E}}^{\mathrm{max}}$, and $\sigma_{t,\mathrm{p}}/\sigma_{t,\mathrm{p}}^{\mathrm{min}}$ for the r.m.s.~electron bunch lengths of $\rho\omega\sigma_{t,\mathrm{e}} = 2$, $4$, and $8$ (blue, red, and green).}
	\label{fig:FEL_3}
	\end{figure}

	The amplification process in a SASE FEL starts from shot noise in the electron beam, passes the stage of exponential amplification ($s/L_{\mathrm{sat}}\approx0.8$), and finally enters the saturation regime ($s/L_{\mathrm{sat}}\approx1$), where $s/L_{\mathrm{sat}}$ is the normalized undulator length with $L_{\mathrm{sat}}$ being the saturation length (see Fig.~\ref{fig:FEL_3}). The FEL radiation pulse energy exhibits shot-to-shot fluctuations, which are larger for shorter pulse durations. The maximum fluctuations are present at the end of the high-gain exponential gain regime. Radiation from a SASE FEL operating in the exponential gain regime possesses the properties of completely chaotic polarized light~\cite{stat-oc,book}. One consequence is that the probability distribution of the energy in the radiation pulse is given by the gamma distribution
	\begin{equation}
	p(E) = \frac{M^M}{\Gamma (M)}
	\left( \frac{E}{\langle E\rangle }\right)^{M-1} \frac{1}{\langle E\rangle }
	\exp \left( -M \frac{E}{\langle E\rangle } \right) \,,
	\label{gamma}
	\end{equation}

	\noindent where $\Gamma (M)$ is the gamma function, $M = 1/(\sigma_{E}/\langle E \rangle)^{2}$, and $\sigma_E = \sqrt{\langle (E -\langle E \rangle )^2 \rangle}$ is the FEL radiation pulse energy spread. The parameter $M$ can be interpreted as the average number of ``degrees of freedom'' or ``modes'' in the radiation pulse (see, e.g., Refs.~\cite{stat-oc,book}). With knowledge of the coherence time $\tau _{\mathrm{c}}$, the FEL radiation pulse duration can be estimated as $\sim M \times \tau _{\mathrm{c}}$. This estimate assumes a high degree of transverse coherence, which is true for the parameter space of well designed SASE FELs like FLASH. In the framework of a one-dimensional model, the maximum value of the 
	coherence time 
	\begin{equation}
	\tau_{\mathrm{c}}^{\mathrm{max}} \simeq  \frac{1}{\rho\omega}
	\sqrt{ \frac{\pi \ln N_{\mathrm{c}} }{18} } 
	\label{tauc}
	\end{equation}
	and the saturation length 
	\begin{equation}
	L_{\mathrm{sat}} \simeq
	\frac{\lambda_{\mathrm{u}}}{4\pi \rho } \left(
	3+ \frac{\ln N_{\mathrm{c}}}{\sqrt{3} } \right) 
	\label{lsat}
	\end{equation}
	\noindent are expressed in terms of the FEL parameter $\rho$~\cite{boni} and the number of the electrons \mbox{$N_{\mathrm{c}} = I/(e\rho\omega)$} cooperating to the SASE FEL process~\cite{stat-oc,book,boni-94}. Here, $\omega/(2\pi) =  c/\lambda $ is the frequency of the amplified wave, $I$ is the electron bunch current, and $\lambda_{\mathrm{u}}$ is the undulator period. A suitable estimate for the parameter $\rho$ comes from the observation that in the parameter range of SASE FELs operating in the extreme-ultraviolet and X-ray wavelength range, the number of field gain lengths to saturation is about 10 (e.g., Ref.~\cite{stat-oc}). The FEL parameter $\rho$ and the coherence time $\tau_{\mathrm{c}}$ are related to the saturation length $L_{\mathrm{sat}}$ as:
	\begin{equation}
	\rho \simeq \lambda_{\mathrm{u}}/L_{\mathrm{sat}} \,,
	\qquad	\tau_{\mathrm{c}}
	\simeq \lambda L_{\mathrm{sat}}/(2\sqrt{\pi}c\lambda_{\mathrm{u}}) \,.
	\label{eq:rho-practical}
	\end{equation}
	These simple physical considerations are confirmed with numerical simulations using the time-dependent simulation code FAST~\cite{fast}. Here, we consider the model of a Gaussian longitudinal electron bunch profile, and trace FEL parameters for different values of the r.m.s.~electron bunch length $\sigma_{t,\mathrm{e}}$.
	Figure~\ref{fig:FEL_3_a} shows the evolution of the energy $E$ in the radiation pulse, the fluctuations of the radiation pulse energy $\sigma_E/ \langle E  \rangle$, and the normalized r.m.s.~FEL radiation pulse duration $\rho\omega\sigma_{t,\mathrm{p}}$ along the undulator. Qualitative observations are that the radiation pulse energy and pulse duration grow with increasing electron bunch length, and the maximum fluctuations of the radiation pulse energy and the minimum r.m.s.~radiation pulse duration are observed at the end of the exponential gain regime ($s/L_{\mathrm{sat}}\lesssim 0.8$). The normalized values of these parameters, i.e., $E/E_{\mathrm{sat}}$, $\sigma _{\mathrm{E}}/\sigma_{\mathrm{E}}^{\max}$, and $\sigma_{t,\mathrm{p}}/\sigma_{t,\mathrm{p}}^{\min}$, exhibit nearly universal dependencies for a normalized electron bunch length $\rho\omega\sigma _{t,\mathrm{e}} \gtrsim 1$ as it is shown in Fig.~\ref{fig:FEL_3_b}.  
	\begin{figure}[t]
	\includegraphics[width=0.9\linewidth]{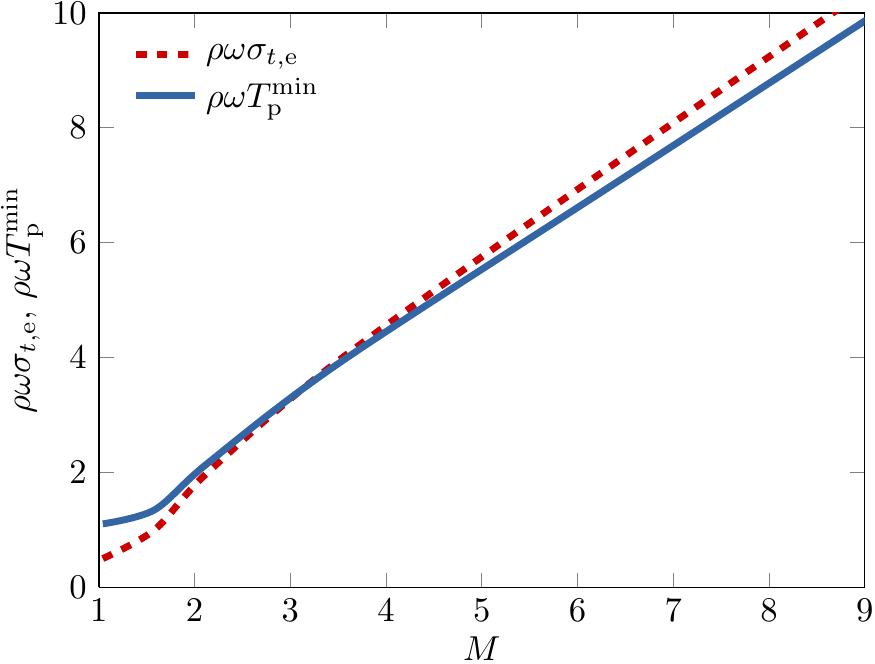}
	\caption{Normalized r.m.s.~electron bunch length $\rho\omega\sigma_{t,\mathrm{e}}$ contributing to the FEL process (red dashed line) and the corresponding normalized minimum FWHM radiation pulse duration $\rho\omega T_{\mathrm{p}}^{\min}$ (blue solid line) at the end of the exponential gain regime ($s/L_{\mathrm{sat}}\lesssim 0.8$) versus the number of modes $M$.}
	\label{fig:FEL_4}
	\end{figure} 
	This allows to derive a universal dependency between the r.m.s.~electron bunch length $\sigma_{t,\mathrm{e}}$ and the full width at half maximum (FWHM) radiation pulse duration $T_{\mathrm{p}}^{\min} = 2\sqrt{2\ln 2 }\,\sigma_{t,\mathrm{p}}^{\min}$ at the end of the exponential gain regime as a function of the number of modes $M$ in the FEL radiation pulse (see Fig.~\ref{fig:FEL_4}). 
	For $M \gtrsim 2$ and with reasonable practical accuracy we have:	\begin{eqnarray}
	\sigma_{t,\mathrm{e}}  \simeq  2\sqrt{2\ln 2 }\,\sigma_{t,\mathrm{p}}^{\min} = T_{\mathrm{p}}^{\min}\simeq
	\frac{M\lambda }{5c\rho } \simeq
	\frac{M\lambda L_{\mathrm{sat}}}{5c\lambda_{\mathrm{u}}} \ .
	\label{eq:taurad-m}
	\end{eqnarray}

	\noindent
	The minimum FWHM FEL radiation pulse duration expressed in terms of the coherence time (cf. Eq.~(\ref{eq:rho-practical})) is $T_{\mathrm{p}}^{\min} \simeq 0.7 \times M \times \tau_{\mathrm{c}}$ . Lengthening of the FEL radiation pulse occurs when the amplification process enters the saturation regime and happens due to two effects. The first effect is lasing toward saturation in the tails of the electron bunch, and the second effect is radiation pulse lengthening due to slippage effects, which is one radiation wavelength per undulator period. The effect of lasing in the tails gives the relative radiation pulse lengthening as it is illustrated in Fig.~\ref{fig:FEL_3_a}. At saturation ($s/L_{\mathrm{sat}}\approx1$), the pulse lengthening is a factor of about 1.4 with respect to the minimum pulse duration in the exponential gain regime ($s/L_{\mathrm{sat}}\lesssim 0.8$) given by Eq.~(\ref{eq:taurad-m}), and it is increased up to a factor of 2 in the deep nonlinear regime ($s/L_{\mathrm{sat}}\gtrsim 1.2$). It is also seen that the slippage effect is more pronounced for relative lengthening of short pulses.

	Measurements of statistical properties of SASE FELs allow to estimate mean photon pulse durations and the lasing fraction of the FEL driving electron bunches, and this technique has been effectively used at FLASH and LCLS~\cite{FLASHAnature,FLASH1,lcls-fel2010}. First, the FEL process is tuned to the maximum radiation pulse energy, which occurs in the saturation regime and beyond ($s/L_{\mathrm{sat}}\gtrsim1$), and then the FEL radiation pulse energy is recorded at different positions along the undulators by applying orbit kicks in order to suppress lasing. The resulting ``FEL gain curve'' permits the estimation of the saturation length $L_{\mathrm{sat}}$, and the FEL parameter $\rho $ and the coherence time $\tau _{\mathrm{c}}$ can be calculated by using Eq.~(\ref{eq:rho-practical}). Then the FEL process is tuned to the end of the exponential regime at $s/L_{\mathrm{sat}}\lesssim 0.8$ where the FEL radiation power is reduced by a factor of $\sim20$ with respect to the saturation regime (see Fig.~\ref{fig:FEL_3_a}). The fluctuations of the FEL radiation pulse energy reach their maximum value at this point, and the inverse squared value of radiation pulse energy fluctuations gives the number of the modes in the radiation pulse. The electron bunch length and the minimum photon pulse duration at the end of the exponential gain regime are then derived from Eq.~(\ref{eq:taurad-m}). The relative photon pulse lengthening at saturation corresponds to a factor of about 1.4, and in the (deep) nonlinear regime, it depends on the electron bunch length (see Fig.~\ref{fig:FEL_3_b}).

	\section{Experimental Results at FLASH}\label{sec:Results}
	Comparative measurements of soft X-ray pulse durations and electron bunch lengths were carried out for two accelerator and beam settings. In the first step, we measured longitudinal electron bunch profiles determined by both time-domain longitudinal phase measurements and frequency-domain THz spectroscopy for electron bunch charges of 100\,pC and 500\,pC in order to verify the consistency of both methods. In the second step, we compared longitudinal electron bunch profiles determined by longitudinal phase measurements with the soft X-ray pulse durations estimated by SASE FEL radiation pulse energy statistics for electron bunch charges of 150\,pC and 500\,pC. The relevant parameters are given in Table~\ref{tab:spec}.
	\begin{table}[t]
	\centering
	\caption{Parameters for the comparative measurements of FEL photon pulse durations and electron bunch lengths. The numbers in the brackets of the electron bunch charge and the FEL radiation saturation level are related to each other.}
	\begin{ruledtabular}
	\begin{tabular} {lcccc}
	Parameter           & Symbol       & Value     & Unit  \\ \hline
	Electron beam energy 	    &$E_{\mathrm{e}}$   & 660  & MeV \\
	FEL radiation wavelength	    &$\lambda$   & 14.6   &nm \\
	Electron bunch charge  &$Q$   & 150 (500)  & pC \\
	Saturation level of FEL radiation &$E_{\mathrm{sat}}$   & 30 (200)   &$\mathrm{\mu J}$ \\
	Undulator period	    &$\lambda_{\mathrm{u}}$   & 27.3   &mm \\
	\end{tabular}
	\label{tab:spec}
	\end{ruledtabular}
	\end{table}

		\subsection{Longitudinal phase space measurements}\label{subsec:Results}
	
\begin{figure}[t]
		\centering
		\subfigure[~$Q=150\,\mathrm{pC}$.]{\includegraphics[width=0.9\linewidth]{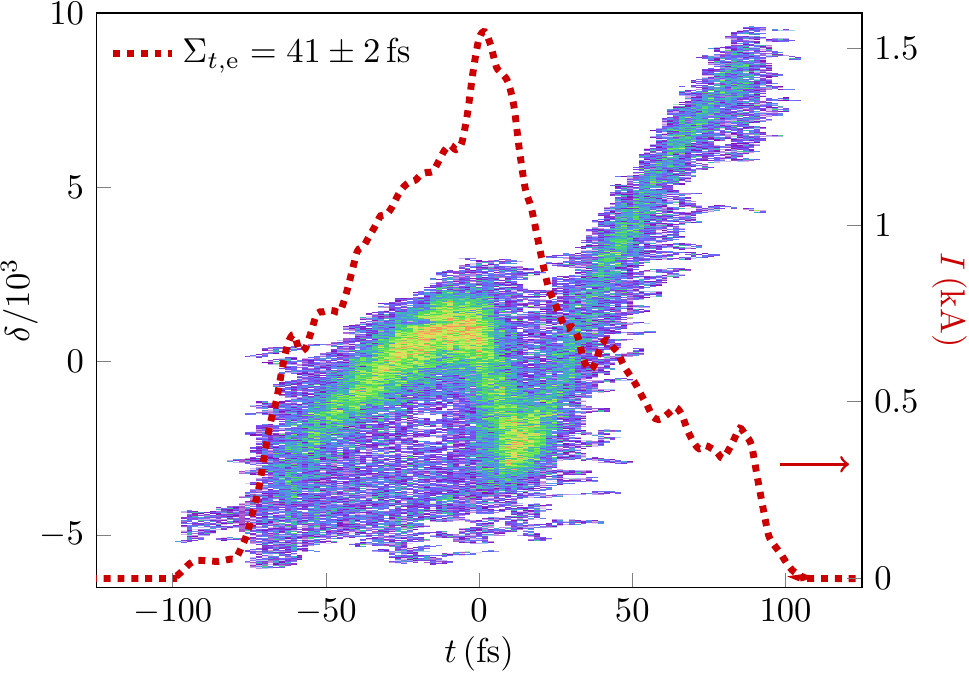} \label{fig:LPS_5_a}}
		\subfigure[~$Q=500\,\mathrm{pC}$.]{\includegraphics[width=0.9\linewidth]{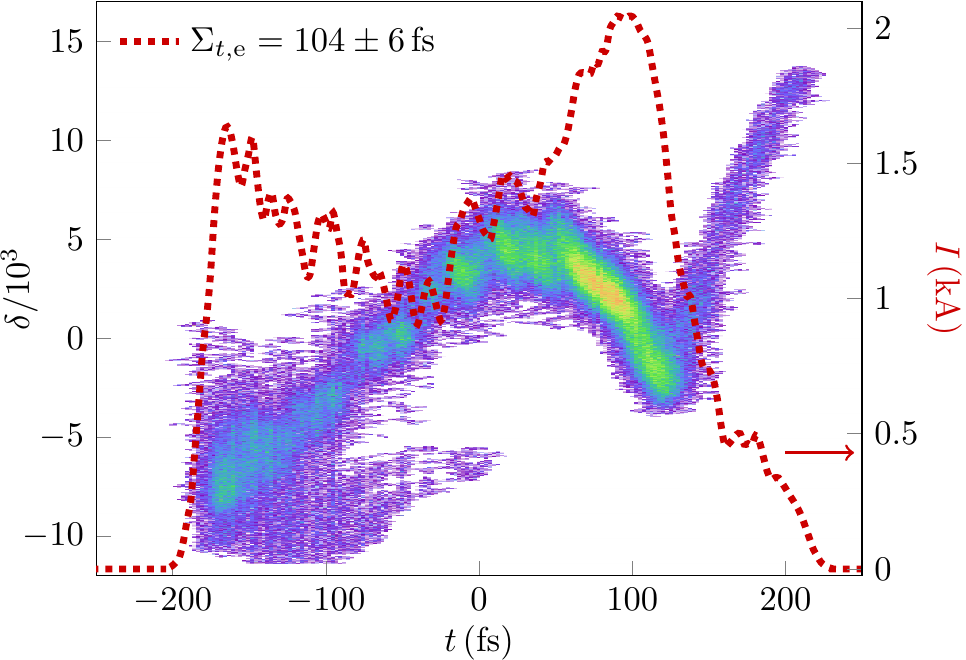} \label{fig:LPS_5_b}}
		\caption{Single-shot longitudinal phase space and electron bunch current (dotted red lines) measurements using the TDS in combination with the energy spectrometer for bunch charges of (a) 150\,pC and (b) 500\,pC. The selected single-shots represent typical bunches out of 50 measurements, and the r.m.s.~bunch length values ($\Sigma_{t,\mathrm{e}}$) including measurement error in the legends correspond to the shown single-shots.}
		\label{fig:LPS_5}
		\end{figure}

		The longitudinal phase space measurements were performed using a $100\,\mathrm{\mu m}$ thick scintillator (YAG:Ce) imaging screen, and the camera system was set up with a spatial resolution of better than $16\,\mathrm{\mu m}$. The time calibration was achieved by measuring the vertical electron bunch centroid on the imaging screen during a TDS rf phase scan. By measuring the intrinsic vertical r.m.s.~beam size, i.e., with the TDS rf power switched off, the r.m.s.~time resolutions was estimated (see Sec.~\ref{subsec:LOLA}) for each measurement individually. For comparisons with soft X-ray pulse durations (see Sec.~\ref{sec:Comparisons}), 50 consecutive single-shot measurements of the longitudinal phase space were taken for each set of electron beam parameters with bunch charges of 150\,pC and 500\,pC. The longitudinal phase space and electron bunch current of typical single-shots, together with the single-shot r.m.s.~bunch lengths, are shown in Fig.~\ref{fig:LPS_5}. The presented measurements were performed for one zero-crossing TDS rf phase, and corrections of a potential longitudinal-to-vertical correlation $C_y$ were neglected with $S_y \gg C_y\approx0$. This is justified by the large shear parameter (see below) and the agreement in the comparative measurements between TDS and THz spectrometer presented in Fig.~\ref{fig:THz_6}, where the latter is not sensitive to longitudinal-to-vertical correlations. 

		The single-shot r.m.s.~time resolution $\mathcal{R}_{t,\mathrm{e}}$ reached unprecedented 8\,fs for the 150\,pC case, which corresponds to a shear parameter of $S_y c^{-1}=30$ (significantly larger than any longitudinal-to-vertical correlation $C_y$ observed at FLASH) and an intrinsic vertical r.m.s.~beam size of $\sigma_{y0}=70\,\mathrm{\mu m}$. For the 500\,pC settings, we achieved an r.m.s.~time resolutions of 13\,fs, which is due to a larger intrinsic vertical beam size. The average r.m.s.~electron bunch lengths, including statistical errors due to fluctuations of electron beam parameters and accelerator settings, of 50 consecutive single-shots are $\langle \Sigma_{t,\mathrm{e}} \rangle=41\pm3\,\mathrm{fs}$ and $103\pm4\,\mathrm{fs}$ for the bunch charges of 150\,pC and 500\,pC, respectively. In order to take account of the unknown correlation $C_y$, we admit a systematical uncertainty of $\pm10\,\%$ in the r.m.s.~electron bunch length measurements.

		The longitudinal phase space has a similar shape in both cases (Figs.~\ref{fig:LPS_5_a} and~\ref{fig:LPS_5_b}) with an overall negative energy chirp $-d\delta/dt$, i.e., electrons in the leading part of the bunch (on the left in Fig.~\ref{fig:LPS_5}) have less energy than those in the trailing part. However, both distributions show a distinct core region with a positive energy chirp, which is most probably generated by space charge forces. In general, the implementation of the new electron bunch compression scheme using the third third-harmonic rf linearizer system~\cite{lin1,lin2} results in electron bunches with significantly more confined bunch current profiles without trailing tails of picosecond duration~\cite{FLASHAnature, FLASH1}.
		
		\subsection{Measurement of longitudinal form factors}\label{subsec:ctr}
		The principle of longitudinal electron beam diagnostics based on spectroscopy of coherent radiation has been described in Sec.~\ref{subsec:kk}. The measurements of the longitudinal form factors were performed by spectroscopy of coherent transition radiation generated from an aluminum coated silicon screen and using the THz spectrometer upstream of the collimator (see Fig.~\ref{fig:FLASH_1}). Technical details about the THz spectrometer and the transport beamline for THz radiation can be found in Refs.~\cite{THz,Wesch1,THZB}. Several hundred single-shot THz spectra in the wavelength range 5-430\,$\mathrm{\mu m}$ were recorded for electron bunch charges of 100\,pC and 500\,pC. At the same time, single-shot longitudinal phase space measurements using the TDS in combination with the energy spectrometer were carried out in order to compare and verify the reconstruction of the longitudinal electron bunch profiles from the spectroscopic measurements. Diagnostics for FEL photon pulse durations were not available during these measurements.
		
		\begin{figure}[b]
			\centering
			\subfigure[~$Q=100\,\mathrm{pC}$.]{\includegraphics[width=0.9\linewidth]{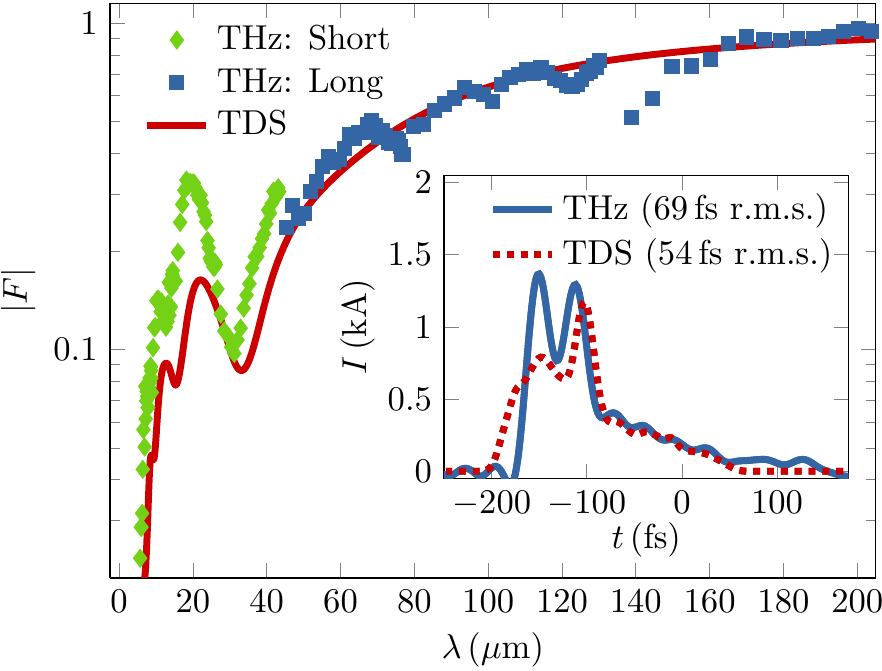} \label{fig:THz_6_a}}
			\subfigure[~$Q=500\,\mathrm{pC}$.]{\includegraphics[width=0.9\linewidth]{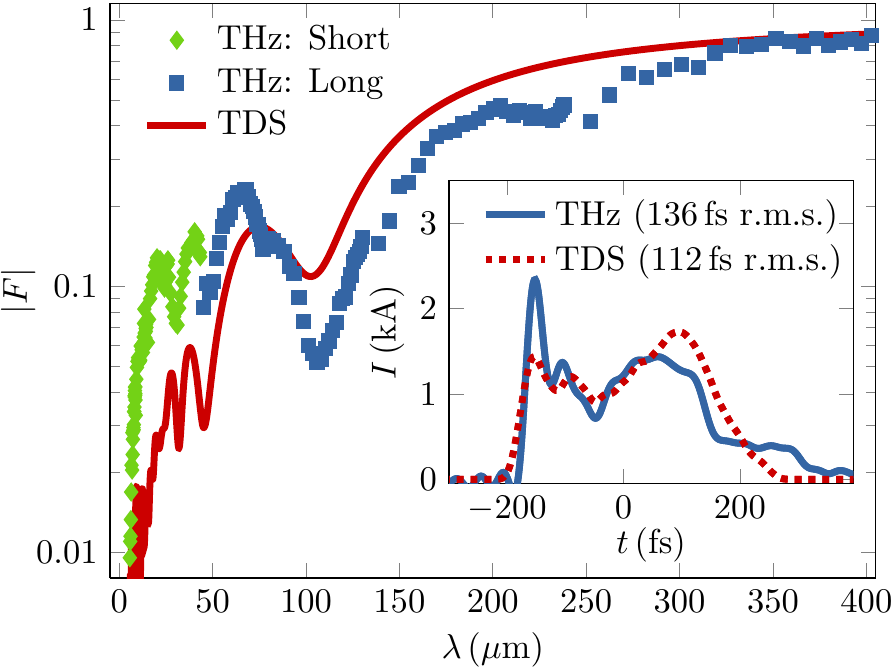} \label{fig:THz_6_b}}
			\caption{Mean moduli of the longitudinal form factors measured by THz spectroscopy of coherent transition radiation (green diamonds and blue squares) and derived from TDS measurements (solid red lines), generated by electron bunches with charges of (a) 100\,pC and (b) 500\,pC, respectively. The insets show single-shot electron bunch currents reconstructed from THz spectra (solid blue lines) in comparison with single-shot TDS measurements (dotted red lines), and the numbers in the brackets represent the calculated r.m.s.~bunch lengths.}
			\label{fig:THz_6}
		\end{figure} 

		Figure~\ref{fig:THz_6} shows the measured mean moduli of the longitudinal form factors, including $|F|$ derived from TDS measurements (red line), together with the reconstructed electron bunch currents in comparison with the corresponding time-domain TDS measurements (insets of Fig.~\ref{fig:THz_6}). The measurements of the longitudinal form factors $|F|$ were done with two reflection grating sets (``THz: Short'' and ``THz: Long'') and averaged over $\sim300$ single shots. The insets of Fig.~\ref{fig:THz_6} show single-shot electron bunch currents reconstructed by means of an inverse Fourier transform using the Kramers-Kronig relations for the phase retrieval in comparison with direct measurements in the time-domain using the TDS. The integrals in Eqs.~(\ref{eq:kk2}) and~(\ref{eq:kk3}) for the electron bunch profile reconstruction were computed for single shots, representing the typical form factor out of the $\sim300$ measurements, with a short-wavelength cut-off at $5\,\mathrm{\mu m}$ and by applying an extrapolation with $|F|\rightarrow 1$ for long wavelengths.				

		The results of both experimental methods are in good agreement and show the same features within the electron bunch currents. The deviations can be explained by uncertainties of both methods, including the unknown and omitted Blaschke phase in Eq.~(\ref{eq:kk2}), and due to the fact that both diagnostics are not located at the same position along the beamline. In fact, the collimator between the two positions (see Fig.~\ref{fig:FLASH_1}) generates longitudinal dispersion, which changes the longitudinal electron bunch profile depending on the actual energy chirp of the electron bunch and the settings of the collimator magnets. Nevertheless, the good overall agreement of the comparative measurement confirms the consistency of both high-resolution longitudinal electron beam diagnostics.

		\subsection{Measurement of FEL pulse energy statistics}\label{subsec:stat}
		The physical background of the FEL photon pulse duration estimations based on statistical properties of SASE FEL radiation has been described in Sec.~\ref{subsec:FEL}. FLASH is equipped with the following set of detectors for single-shot measurements of the radiation pulse energy: gas monitor detectors, micro-channel plate (MCP) based detectors, photodiodes, and thermopiles~\cite{gmd-detector,mcp-detector,mcp-detector-2007,tiedtke-njp-2009}. The detectors are installed at several positions along the photon beamline (see also Fig.~\ref{fig:FLASH_1}). The MCP detector is installed in front of all other detectors and is used for precise measurements of the FEL photon pulse energy. The MCP measures the radiation scattered by a metallic mesh (Cu, Fe, and Au targets are being used) placed behind an aperture located downstream of the undulator magnets. The electronics of the MCP detector has an excellent signal-to-noise ratio of about 100 and thus allows high-resolution measurements of the energy fluctuation. 

		The measurement procedure was organized as follows. First, the SASE FEL process was tuned to the maximum MCP signal at full undulator length of 27\,m (six undulator modules). Then an orbit kick was applied by switching on steerer magnets after the fourth undulator module such that the FEL amplification process was suppressed in the last two undulator modules. The level of the FEL radiation pulse energy after four undulator modules was about a factor of $\sim20$ less compared to the level at saturation, and by taking into account the saturation length, which was estimated to be $L_{\mathrm{sat}}=22.5\pm2.5\,\mathrm{m}$ for both electron bunch charges within the presented experimental conditions, the normalized undulator length was $s/L_{\mathrm{sat}}=0.80\pm0.09$ there. It has been shown in Sec.~\ref{subsec:FEL} that this point corresponds to the end of the exponential gain regime. The aperture of the detector was adjusted such that all photons passed through the scattering target of the detector. Thus, the MCP signal is proportional to the single-shot FEL radiation pulse energy. However, the fluctuations of electron beam and accelerator parameters contribute to the fluctuations of the radiation pulse energies, while only fundamental SASE FEL fluctuations are essential. For this reason, a selection procedure had to be applied to the recorded data. The important electron beam and accelerator parameters, e.g., bunch charge monitor readings, beam position monitor readings, readings of the bunch compression monitors, and read back values of rf parameters were recorded for each shot together with the readings of the MCP detector. If electron beam or accelerator parameters deviated more than the prescribed threshold, the events were excluded from the data set. We note that the number of events in the data set after the selection procedure must be sufficiently large in order to provide an acceptable statistical accuracy. 
		\begin{figure}[t]
			\centering
			\subfigure[~$Q=150\,\mathrm{pC}$.]{\includegraphics[width=0.9\linewidth]{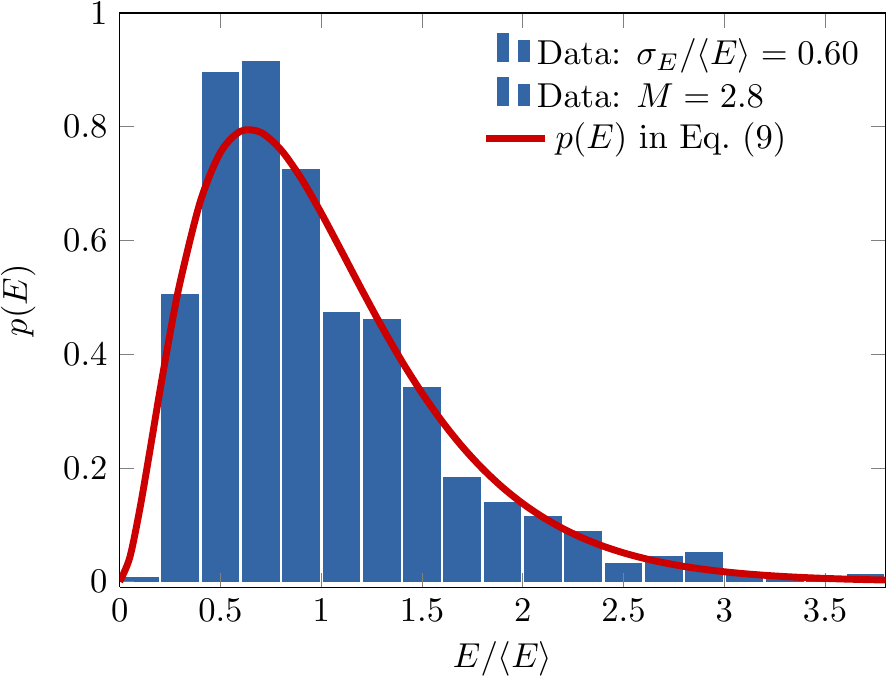} \label{fig:Fig_Stat_7_a}}
			\subfigure[~$Q=500\,\mathrm{pC}$.]{\includegraphics[width=0.9\linewidth]{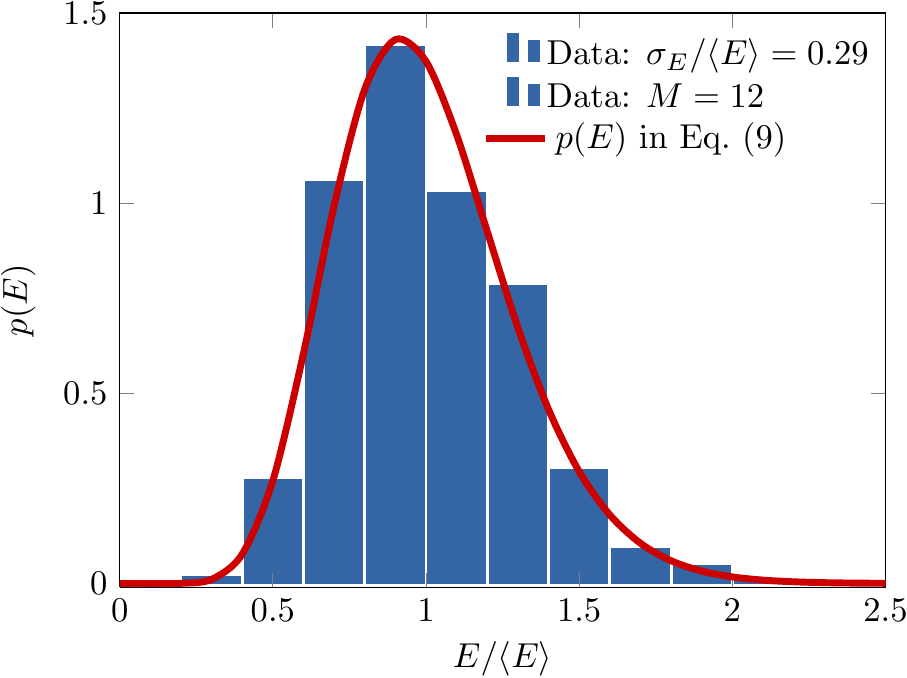} \label{fig:Fig_Stat_7_b}}
			\caption{The probability distributions $p(E)$ of the FEL radiation pulse energy $E$ of soft X-ray pulses at the end of the exponential gain regime for electron bunch charges of (a) 150\,pC and (b) 500\,pC. The histograms (blue bars) represent the data with the relative radiation pulse energy spread $\sigma_E/\langle E  \rangle$, and the solid curve (red lines) show the gamma distribution using $M = 2.8$ and $12$ (modes) for 150\,pC and 500\,pC, respectively.}
		\label{fig:Stat_7}
		\end{figure}
		
		\begin{figure}[htb]
			\centering
			\subfigure[~$Q=150\,\mathrm{pC}$.]{\includegraphics[width=0.9\linewidth]{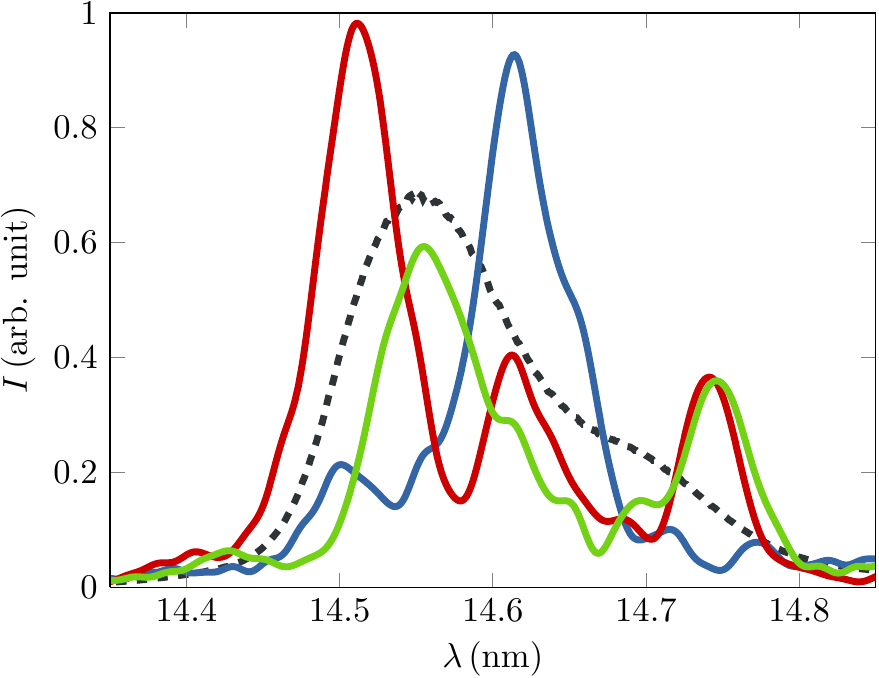} \label{fig:Fig_Spec_8_a}}
			\subfigure[~$Q=500\,\mathrm{pC}$.]{\includegraphics[width=0.9\linewidth]{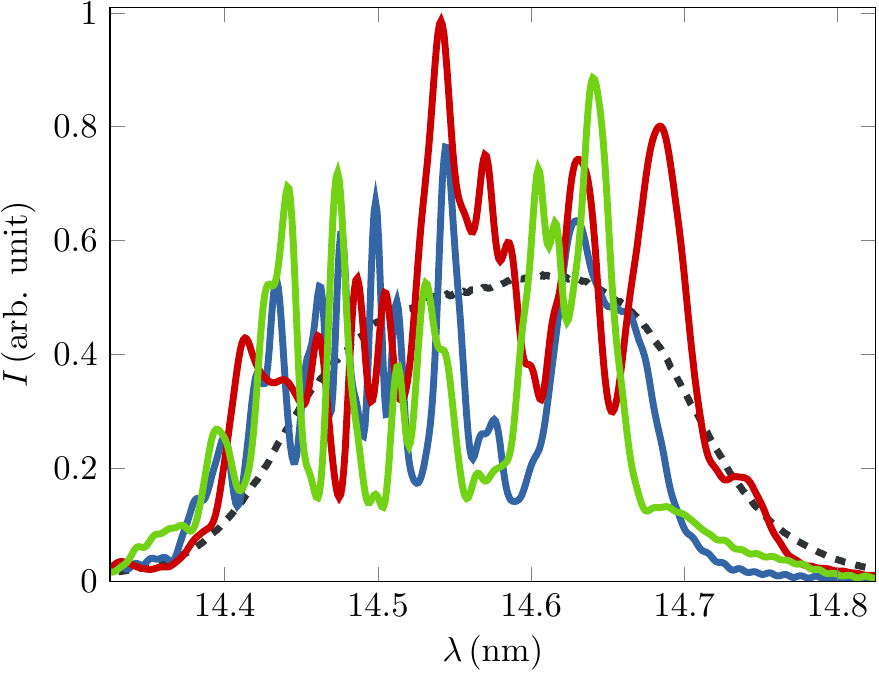} \label{fig:Fig_Spec_8_b}}
			\caption{Single-shot (red, green, and blue solid lines) and mean (dashed black lines) spectra of soft X-ray pulses  measured at the end of the exponential gain regime using the SXR spectrometer for the settings with (a) 150\,pC and (b) 500\,pC. The mean spectra represent an average over $\sim 2500$ single shots.}
		\label{fig:Spec_8}
		\end{figure}

		Figure~\ref{fig:Stat_7} shows the probability distribution of the FEL radiation	pulse energies for the electron bunch charges of 150\,pC and 500\,pC. The sets of raw data contained about 1400 (1700) single-shot measurements, and about 800 (550) measurements remained after the data selection procedure for 150\,pC (500\,pC). The selection was mainly achieved by removing large electron bunch charge fluctuations and rf phase drifts in the first accelerating structure (ACC1) upstream of the first bunch compressor. The data selection thresholds were chosen to remove clear correlations with the MCP readings. The measured number of modes amount to $M=2.8\pm0.5$ and $12\pm2$ for the settings with 150\,pC and 500\,pC, respectively, and the quoted errors are based on uncertainties of the data selection procedure. If a Gaussian shape for the lasing fraction of the electron bunch is assumed, we can apply Eq.~(\ref{eq:taurad-m}) in order to determine the r.m.s.~electron bunch length that contributes to lasing and get $\sigma_{t,\mathrm{e}}=23\pm4\,\mathrm{fs}$ and $96\pm18\,\mathrm{fs}$ for the electron bunch charge of 150\,pC and 500\,pC, respectively. The minimum FWHM photon pulse durations $T_{\mathrm{p}}^{\min}$ at the end of the exponential gain regime are estimated to be these obtained r.m.s.~electron bunch lengths (see Eq.~(\ref{eq:taurad-m})), i.e., $T_{\mathrm{p}}^{\min} =23\pm4\,\mathrm{fs}$ and $96\pm18\,\mathrm{fs}$, respectively. The quoted errors are purely statistical and assume a Gaussian shape for the lasing fraction of the electron bunch. Significant systematical errors can appear due to large deviations from the assumed longitudinal electron bunch profile and time-dependent electron bunch properties such as the transverse slice emittance.

		In parallel with statistical measurements, FEL photon spectra were recorded using the soft X-ray (SXR) spectrometer in the photon beamline. We note that the spectrum is simply a Fourier transform of the temporal structure, and the average number of spikes (modes) in the time-domain should be about the same as the number of spikes in the spectral-domain. The single-shot soft X-ray spectra were obtained by using a plane grating monochromator, operating in a spectrographic mode with a resolving power $\lambda/\Delta \lambda > 10000$~\cite{pgm}. We observed qualitative agreement between the spectral measurements (number of spikes) and the number of modes measured by FEL pulse energy statistics (see Fig.~\ref{fig:Spec_8}). The detailed analysis of the spectral measurements, taking into account the energy chirp of the electron bunches, is beyond the scope of this paper and will be reported elsewhere. 

		Based on the theoretical model described in Sec.~\ref{subsec:FEL} and according to Fig.~\ref{fig:FEL_3_b}, the expected FEL photon pulse durations at saturation and beyond can be extrapolated by the minimum FWHM photon pulse durations measured at the end of the exponential gain regime. In the nonlinear regime at $s/L_{\mathrm{sat}}\approx 1.2$ (see Fig.~\ref{fig:FEL_3_b}) and by using the measured minimum FWHM photon pulse durations $T_{\mathrm{p}}^{\min}$, we would expect FWHM FEL photon pulse durations of $T_{\mathrm{p}}=43\pm8\,\mathrm{fs}$ (lengthening of $\sim1.9$) and $T_{\mathrm{p}}=169\pm32\,\mathrm{fs}$ (lengthening of $\sim1.75$) for the beam settings with 150\,pC and 500\,pC, respectively. For these expected FEL photon pulse durations and by considering the measured average FEL pulse energies (see Table~\ref{tab:spec}) of $30\,\mathrm{\mu J}$ (150\,pC) and $200\,\mathrm{\mu J}$ (500\,pC), the peak powers result in $0.7\pm0.1\,\mathrm{GW}$ and $1.1\pm0.2\,\mathrm{GW}$, respectively.

	\section{Soft X-ray pulse durations versus electron bunch lengths}\label{sec:Comparisons}
	The average r.m.s.~electron bunch lengths determined by time-domain longitudinal phase space diagnostics are $\langle \Sigma_{t,\mathrm{e}} \rangle=41\pm3\,\mathrm{fs}$ and $103\pm4\,\mathrm{fs}$ for the bunch charges of 150\,pC and 500\,pC, respectively. The corresponding minimum FWHM soft X-ray pulse durations at the end of the exponential gain regime estimated by SASE FEL pulse energy statistics are $T_{\mathrm{p}}^{\min} = 23\pm4\,\mathrm{fs}$ and $96\pm18\,\mathrm{fs}$, respectively for the same electron bunch charges. 
	
\begin{figure}[t]
	\includegraphics[width=0.9\linewidth]{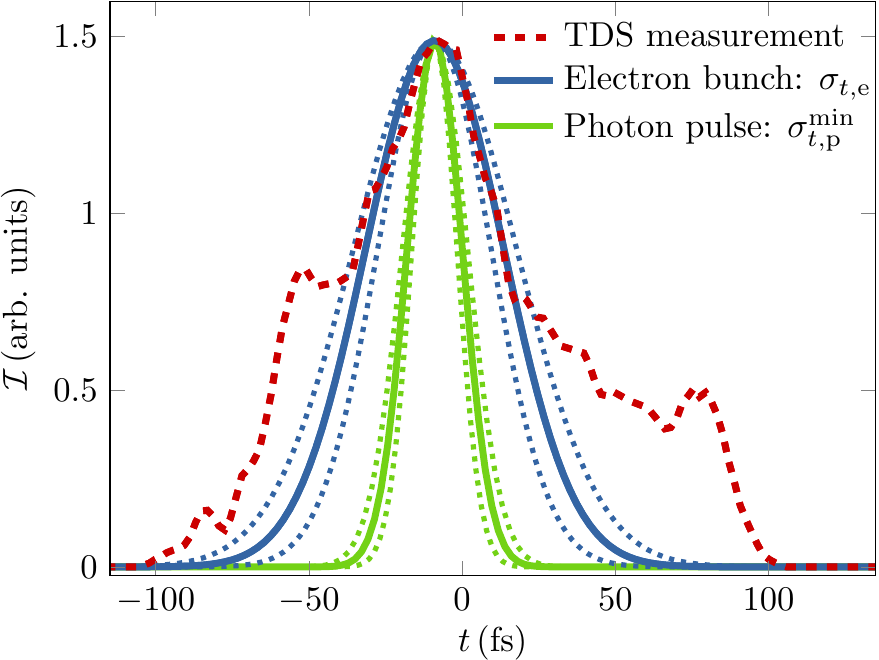}
	\caption{Single-shot longitudinal electron bunch profile measurement using the TDS for a bunch charge of 150\,pC (dashed red line). The selected single-shot represents a typical bunch out of 50 measurements (cf. Fig.~\ref{fig:LPS_5_a}) with an r.m.s.~bunch length of $\Sigma_{t,\mathrm{e}} = 41\pm2\,\mathrm{fs}$. The blue solid (dotted) line shows a Gaussian with $\sigma= 23\pm4\,\mathrm{fs}$ representing the width $\sigma_{t,\mathrm{e}}$ of a potential lasing part of the electron bunch. The green solid (dotted) line shows a Gaussian with $\sigma= (23\pm4)/(2\sqrt{2 \ln 2})\,\mathrm{fs}$ representing the corresponding soft X-ray (photon) pulse duration $\sigma_{t,\mathrm{p}}^{\min}$. The dotted lines describe the error bands ($\pm$).}
	\label{fig:FEL_8}
	\end{figure}

	As described in Sec.~\ref{subsec:FEL} for Gaussian longitudinal profiles, the r.m.s.~electron bunch length $\sigma_{t,\mathrm{e}}$ that contributes to lasing is related to the minimum FWHM FEL photon pulse duration $T_{\mathrm{p}}^{\min}$ by $\sigma_{t,\mathrm{e}}  \simeq T_{\mathrm{p}}^{\min}$ (see Eq.~(\ref{eq:taurad-m})), where $T_{\mathrm{p}}^{\min} = 2\sqrt{2\ln 2 }\,\sigma_{t,\mathrm{p}}^{\min}$ with the corresponding minimum r.m.s.~FEL photon pulse duration $\sigma_{t,\mathrm{p}}^{\min}$. The comparison of $\Sigma_{t,\mathrm{e}}$ and $T_{\mathrm{p}}^{\min}$ ($\sigma_{t,\mathrm{e}}$) shows that the r.m.s.~electron bunch lengths measured by longitudinal phase space diagnostics are larger than estimated by SASE FEL pulse energy statistics, which is more significant for the bunch charge of 150\,pC. Figure~\ref{fig:FEL_8} shows a measured single-shot longitudinal electron bunch profile ($I$ in kA) for 150\,pC and two Gaussians ($\mathcal{I}$ in arb. units) with standard deviations $\sigma=\sigma_{t,\mathrm{e}}  \simeq T_{\mathrm{p}}^{\min}$, which represents a potential lasing part of the electron bunch, and $\sigma=\sigma_{t,\mathrm{p}}^{\min}=T_{\mathrm{p}}^{\min}/(2\sqrt{2\ln 2})$, representing the corresponding soft X-ray pulse. The amplitudes and peak positions of the Gaussians were chosen to fit the core region of the measured longitudinal electron bunch profile. The deviations of the estimated minimum FWHM soft X-ray pulse duration $T_{\mathrm{p}}^{\min}$, i.e., the lasing fraction of the electron bunch $\sigma_{t,\mathrm{e}}$, from the measured r.m.s.~electron bunch length $\Sigma_{t,\mathrm{e}}$ can be explained most likely by the non-Gaussian longitudinal electron beam profile with low currents in the tails that do not contribute to lasing but to the r.m.s.~electron bunch length value.

	In cases of non-Gaussian longitudinal electron bunch profiles, e.g., an electron bunch compression mode with a short leading spike (e.g., Refs.~\cite{FLASHAnature, FLASH1}), a Gaussian-like core region needs to be identified in order to determine the pulse durations. For instance, the FWHM bunch length value, which is not that sensitive to tails in the electron bunch, can be determined and used to calculate an effective r.m.s.~bunch length for comparison with the minimum FWHM FEL photon pulse duration. For the settings with 150\,pC, the FWHM electron bunch length is $70\pm17\,\mathrm{fs}$ which corresponds to an effective r.m.s.~electron bunch length of $30\pm7\,\mathrm{fs}$ (cf. with $T_{\mathrm{p}}^{\min} = 23\pm4\,\mathrm{fs}$). 

	For a more precise identification of the lasing part within the electron bunches, simultaneous information of other parameters, e.g., the slice emittance, or time-domain measurements of the FEL photon pulse profiles (e.g., Refs.~\cite{THZ_streak1,THZ_streak2,xtcav}) are required. However, the longitudinal electron beam diagnostics discussed here provide robust estimates on the upper limits of the expected FEL photon pulse durations and allows their monitoring.

	\section{Summary and conclusions}\label{sec:Summary}
	Information on the FEL photon pulse durations is important for time-resolved and high intensity experiments in photon science. However, direct measurement of femtosecond FEL photon pulse durations in the extreme-ultraviolet and X-ray wavelength range is a tremendous challenge, and well established methods are not yet available. The theoretical considerations, simulations, and experimental data presented in this paper show that high-resolution electron beam diagnostics can be utilized to provide reasonable constraints on expected FEL photon pulse durations. We discussed two different longitudinal electron beam diagnostics with single-shot capability and femtosecond accuracy, and presented consistent single-shot measurements of longitudinal electron bunch profiles with good agreement. The electron beam measurements were performed in the frequency-domain by THz spectroscopy of coherent transition radiation and in the time-domain by longitudinal phase space measurements using a TDS in combination with a magnetic energy spectrometer. In the latter case, an unprecedented single-shot r.m.s.~time resolution of 8\,fs has been achieved. The results from the longitudinal phase space measurements were compared with soft X-ray pulse durations estimated by SASE FEL pulse energy statistics, and the consistent results show that r.m.s.~electron bunch lengths, determined by longitudinal electron beam diagnostics, provide upper limits on the expected FWHM FEL photon pulse durations. Theoretical considerations and FEL simulations verify this experimental observation and facilitate the extrapolation of the expected FEL photon pulse durations in the saturation regime and beyond based on the measurements in exponential gain regime. In addition, we demonstrated the generation of soft X-ray pulses with a peak radiation power in the GW level and with durations well below 50\,fs (FWHM) after implementation of the new uniform electron bunch compression scheme using a third-harmonic rf linearizer system at FLASH.
	
	\begin{acknowledgments}
	We would like to thank the entire FLASH-team, and the engineers and technicians of the DESY groups FLA, MCS, and MVS for their support. In particular, we thank T. Limberg, S. Schreiber, and R. Treusch for general support, B. Faatz and K. Honkavaara for providing beam time, and P. Emma and Z. Huang for helpful discussions.
	\end{acknowledgments}

\end{document}